\documentclass[english,aps,pra,showpacs,superscriptaddress,reprint,twocolumns]{revtex4-1}
\usepackage[english]{babel}
\usepackage[T1]{fontenc}
\usepackage{amsmath,amssymb,graphicx,mathrsfs,xcolor,babel,textcomp,esint}
\usepackage[unicode=true,pdfusetitle,bookmarks=true,bookmarksnumbered=false,bookmarksopen=false,breaklinks=true,pdfborder={0 0 0},backref=false,colorlinks=true]{hyperref}

\begin{document}

\title{Terahertz field induced near-cutoff even-order harmonics in femtosecond
laser}

\author{Bing-Yu Li}
\affiliation{Shanghai University, Shanghai 200444, China}
\affiliation{Shanghai Advanced Research Institute, Chinese Academy of Sciences, Shanghai 201210, China}
\author{Yizhu Zhang}
\email{zhangyz@sari.ac.cn}
\affiliation{Center for Terahertz waves and College of Precision Instrument and Optoelectronics Engineering, Key Laboratory of Opto-electronics Information and Technical Science, Ministry of Education, Tianjin University, Tianjin 300350, China}
\author{Tian-Min Yan}
\email{yantm@sari.ac.cn}
\affiliation{Shanghai Advanced Research Institute, Chinese Academy of Sciences, Shanghai 201210, China}
\author{Y. H. Jiang}
\email{jiangyh@sari.ac.cn}
\affiliation{Shanghai Advanced Research Institute, Chinese Academy of Sciences, Shanghai 201210, China}
\affiliation{University of Chinese Academy of Sciences, Beijing 100049, China}
\affiliation{ShanghaiTech University, Shanghai 201210, China}

\begin{abstract}
  High-order harmonic generation by femtosecond laser pulse in the presence of
  a moderately strong terahertz (THz) field is studied under the strong field
  approximation, showing a simple proportionality of near-cutoff even-order
  harmonic (NCEH) amplitude to the THz electric field. The formation of the
  THz induced-NCEHs is analytically shown for both continuous wave and
  Gaussian pulse. The perturbation analysis with regard to the frequency ratio
  of the THz field to the femtosecond pulse shows the THz-induced NCEHs
  originates from its first-order correction, and the available parametric
  conditions for the phenomenon is also clarified. As the complete
  characterization of the time-domain waveform of broadband THz field is
  essential for a wide variety of applications, the work provides an
  alternative time-resolved field-detection technique, allowing for a robust
  broadband characterization of pulses in THz spectral range.
\end{abstract}
\maketitle

\section{Introduction}\label{sec:introduction}

The development of terahertz (THz) technology has motivated a broad range of
scientific studies and applications in material science, chemistry and
biology. The THz light is especially featured by the availability to access
low-energy excitations, providing a fine tool to probe and control
quasi-particles and collective excitations in solids, to drive phase
transitions and associated changes in material properties, and to study
rotations and vibrations in molecular systems {\cite{salen_matter_2019}}.

In THz science, the capabilities of ultra-broadband detection are essential
to the diagnostics of THz field of a wide spectral range. The most common
detection schemes are based on the photoconductive switches (PCSs)
{\cite{jepsen_generation_1996,van_exter_characterization_1990}} or the
electro-optic sampling (EOS) {\cite{wu_freespace_1995}}. Especially the EOS
technique, which uses part of the laser pulse generating the THz field to
sample the latter, has been widely applied in the THz time-domain
spectroscopy, pump-probe experiments and dynamic matter manipulation
{\cite{lee_principles_2009}}. Nevertheless, both PCSs and EOS require
particular mediums --- photoconductive antenna for the former and
electro-optic crystal for the latter. Due to inherent limitations of the
detection media, including dispersion, absorption, long carrier lifetime, and
lattice resonances {\cite{ferguson_materials_2002}}, the typical accessible
bandwidth of detected THz is limited below 7 THz
{\cite{shen_ultrabroadband_2003,somma_ultra-broadband_2015,kubler_ultrabroadband_2005,wu_7_1997,zhang_broadband_2016}}.

On the other hand, gas-based schemes, including air-breakdown coherent
detection {\cite{dai_detection_2006}}, air-biased coherent detection (ABCD)
{\cite{karpowicz_coherent_2008}}, optically biased coherent detection
{\cite{li_broadband_2015}}, THz radiation enhanced emission of fluorescence
{\cite{liu_broadband_2010}} {{\em et al.\/}}, allow for ultra-broadband
detections since gases being continuously renewable show no appreciable
dispersion or phonon absorption {\cite{lu_investigation_2014}}, thus
effectively extending the accessible spectral range beyond 10 THz. In
particular, the ABCD utilizes the THz-field-induced second harmonics (TFISH)
{\cite{cook_terahertz-field-induced_1999}} to sense the THz transient through
a third-order nonlinear process. The THz field mixed with a bias electric
field breaks the symmetry of air and induces the frequency doubling of the
propagating probe beam. Such a nonlinear mixing results in a signal of the
intensity proportional to the THz electric field, allowing for the coherent
detection by which both amplitude and phase of the THz transient can be
reconstructed.

The air-based broadband THz detection utilizing the laser induced air plasma
as the sensor medium is essentially an inverse process of THz wave generation
(TWG) in femtosecond laser gas breakdown plasma. Both involve rather
complicated processes dominated by the strong field photoionization. From the
perspective of single-atom based strong field theory, the TWG has been
interpreted as the near-zero-frequency radiation due to the
continuum-continuum transition of photoelectron
{\cite{zhou_terahertz_2009,zhang_continuum_2020,zhang_experimental_2020}},
complementary to the widely acknowledged mechanism for high-order harmonic
generation (HHG) --- the continuum-bound transition that emits high energy
photons when the released electron after the ionization recollides with the
parent ion {\cite{lewenstein_theory_1994}}.

Under the same nonperturbative theoretical framework depicting the
strong-field induced radiation, the TWG and the HHG within one atom, however,
present different facets of dynamics of electron wave packet, providing
potential strategies to either characterize the system or to profile external
light fields. For example, the synchronized measurements on angle-resolved TWG
and HHG from aligned molecules allow for reliable descriptions of molecular
structures {\cite{huang_joint_2015}}. On the other hand, the presence of THz
or static fields may drastically affect the photoionization dynamics and
reshape electron trajectories, eventually altering photoelectron spectra and
HHG radiations. Applying the widely used streaking technique, the influence on
photoelectron spectra from an additional THz transient can help characterize
the time structure of an attosecond pulse train, e.g., pulse duration of
individual harmonic {\cite{ardana-lamas_temporal_2016}}. The altered HHG
spectral features by THz or static fields include the increased odd-order
harmonic intensity in the low end of the plateau, the significant production
of even-order harmonics {\cite{bao_static-electric-field_1996}}, the
double-plateau structure {\cite{wang_effects_1998}} and high-frequency
extension {\cite{taranukhin_high-order_2000,hong_few-cycle_2009}}. The
substantially extended HHG cutoff in combined fields creates ultrabroad
supercontinuum spectrum, allowing for the generation of single attosecond
pulses. For example, the combination of a chirped laser and static electric
field has been proposed to obtain isolated pulse as short as 10 attoseconds
{\cite{xiang_control_2009}}.

In this work, we study the influence of an additional moderately strong THz
field on the HHG by femtosecond laser pulse. By "moderately" we mean the THz
field is easily attainable in nowadays laboratories --- its intensity is not
as high as to modify global harmonic features as considered in previous
theoretical works. Instead, we focus on the more subtle THz-induced even-order
harmonics. The dependence of all-order HHG on time delay between the THz field
and the femtosecond pulse is studied, showing the near-cutoff even-order
harmonics (NCEHs) are of particular synchronicity with the external THz field.
Accordingly, the measurement of THz induced NCEHs, similar to the widely used
TFISH, is supposed to provide an alternative all-optical ultrabroad bandwidth
method to characterize the time-domain THz transient. The influence of the THz
field on NCEHs is theoretically investigated under the strong field
approximation, showing a direct link inbetween, further confirming the
availability of the detection scheme.

The paper is organized as follows: In Sec. \ref{sec:scheme}, we present
time-delay dependent all-order HHG calculations and show the significant
correlation between NCEHs and the external THz field, illustrating the
possible schematics to reconstruct the time-domain THz field using NCEHs. In
Sec. \ref{sec:analysis}, the detailed analysis of the NCEHs is presented. A
brief retrospect of the analytical derivation for odd-order harmonics is given
in Sec. \ref{subsec:mono}, followed by the discussion in Sec.
\ref{subsec:mono-and-THz} on the simplest case, the NCEHs in fields of
continuous wave. When a femtosecond laser has a finite pulse width, the
influence from pulse envelope is shown in Sec. \ref{subsec:gaussian}, where
the action for a Gaussian envelope is explicitly derived. In Sec.
\ref{subsec:gaussian-and-THz}, a complete description of the THz-induced NCEHs
in femtosecond pulse is presented, showing the relation between the NCEH and
the THz field at the center of the femtosecond laser pulse. In the end,
exemplary reconstruction schematics are shown with parametric conditions
discussed for the applicability.

\section{Scheme of Coherent Detection and Numerical
Simulations}\label{sec:scheme}

We consider an atom subject to combined fields including a linearly polarized
femtosecond laser pulse $E_0 (t)$ and a THz field $E_1 (t)$. With both
polarizations along the same direction, the total fields read $E (t) = E_0 (t)
+ E_1 (t)$. Denoting the associated vector potentials by $A (t) = A_0 (t) +
A_1 (t)_{}$, we evaluate the harmonics using the Lewenstein model
{\cite{lewenstein_theory_1994}} under the strong field approximation
{\cite{keldysh_ionization_1965,faisal_multiple_1973,reiss_effect_1980}}, which
is usually applicable in the tunneling regimes, providing a reasonable and
intuitive description of harmonic radiation from highly energetic recolliding
electrons. The time-dependent dipole moment $d (t)$ in Ref.
{\cite{lewenstein_theory_1994}} as an integration over the intermediate
momentum can be dramatically simplified by applying the stationary phase
approximation, yielding
\begin{eqnarray}
  d (t) & = & \text{i} \int_0^{\infty} \mathrm{d} \tau \left(
  \frac{\pi}{\epsilon + \text{i} \tau / 2} \right)^{3 / 2} \mu^{\ast}
  [p_{\ensuremath{\operatorname{st}}} (t, \tau) + A (t)]  \nonumber\\
  &  & \times \mu [p_{\ensuremath{\operatorname{st}}} (t, \tau) + A (t -
  \tau)] E_{} (t - \tau) \mathrm{e}^{- \text{i} S (t, \tau)} + \text{c.c.}, 
  \label{eq:d-sfa-generic}
\end{eqnarray}
where integration variable $\tau$ is the return time of the electron, i.e.,
the interval between the instants of ionization and rescattering. In this
work, atomic units are used unless noted otherwise. The dipole matrix element
$\mu (k) = \langle k | \hat{x} | \Psi_0 \rangle$ between bound state $| \Psi_0
\rangle$ and continuum state $| k \rangle$ of momentum $k$ is given by
$\mathrm{i} \partial_k \langle k| \Psi_0 \rangle$ along the polarization
direction. Taking $| \Psi_0 \rangle$ the $1 s$ state of a hydrogen-like atom
for example, $\mu (k) = - \text{i} 2^{7 / 2}  (2 I_p)^{5 / 4} k / [\pi (k^2 +
2 I_p)^3]$, where $I_p$ is the ionization potential. In Eq.
(\ref{eq:d-sfa-generic}), the action reads
\begin{eqnarray}
  S (t, \tau) & = & \int_{t - \tau}^t \mathrm{d} t'  \left( \frac{1}{2} 
  [p_{\ensuremath{\operatorname{st}}} (t, \tau) - A (t')]^2 + I_p \right), 
  \label{eq:action-S}
\end{eqnarray}
and the stationary momentum $p_{\ensuremath{\operatorname{st}}} (t, \tau) = -
[\alpha (t) - \alpha (t - \tau)] / \tau$ is determined by the electron
excursion $\alpha (t) = \int^t \mathrm{d} t' A (t')$ after the electron is
released by external light fields. The subsequent spread of the continuum
electron wave packet is depicted by $\left[ \pi / \left( \epsilon + \text{i}
\tau / 2 \right) \right]^{3 / 2}$ in the integral of Eq.
(\ref{eq:d-sfa-generic}) with an infinitesimal $\epsilon$.

\begin{figure}[h]
  \resizebox{226pt}{226pt}{\includegraphics[scale=1]{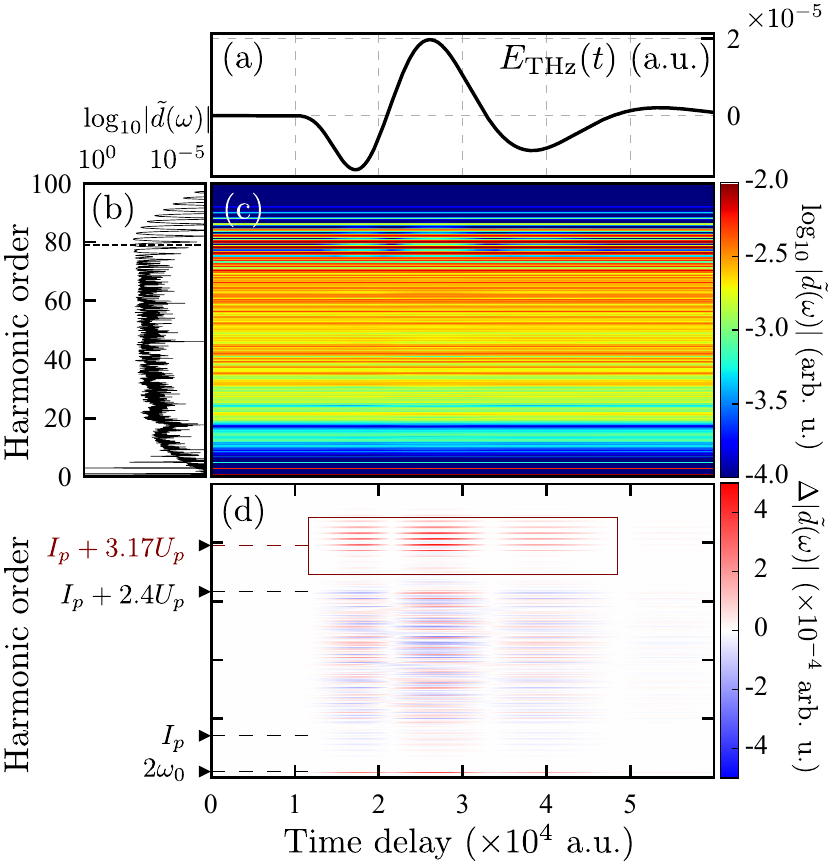}}
  \caption{\label{fig:scan-2d}Schematics of the reconstruction of THz field
  with near-cutoff even-order harmonics, which uses a femtosecond laser pulse
  to scan over a THz field and to record the generated harmonics as a function
  of the time delay between the fields. Panel (a) shows an exemplary THz field
  with $\omega_1 = 2 \times 10^{- 4}$ (1 THz), and $E_1 = 2 \times 10^{- 5}$
  (100 kV/cm). Without the THz field, panel (b) shows the harmonics generated
  by a femtosecond pulse with \ $\omega_0 = 0.0353$ (1.3 $\mu$m), $E_0 = 0.06$
  ($I = 1.3 \times 10^{14}$ W/cm$^2$) and $\sigma = 2106$ (FWHM of 120 fs).
  With the THz field, panel (c) shows the all-order harmonics versus the time
  delay. The difference $\Delta | \tilde{d} (\omega) |$ between the harmonics
  with THz field, as given in (c), and the one without THz field, as shown in
  (b), is presented in panel (d) in the linear scale. All near-cutoff
  even-order harmonics, as indicated by the box around $E_{\text{cutoff}} =
  I_p + 3.17 U_p$, exhibits synchronous change with the intensity of the THz
  field.}
\end{figure}

Evaluating $d (t)$ of Eq. (\ref{eq:d-sfa-generic}) and $| \tilde{d} (\omega)
|$ from its Fourier transform, we demonstrate in Fig. \ref{fig:scan-2d} the
all-order harmonics as a function of the time delay between a near-infrared
pulse and a THz field. The near-infrared pulse has the vector potential of a
Gaussian envelope, $A_0 (t) = (E_0 / \omega_0) \mathrm{e}^{- t^2 / (2
\sigma^2)} \sin (\omega_0 t)$, with $\omega_0 = 0.0353$ (1.3 $\mu$m), $E_0 =
0.06$ (intensity $I = 1.3 \times 10^{14}$ W/cm$^2$), $\sigma = 2106$ (FWHM of
120 fs). The THz field is modeled by $A_1 (t) \propto - (E_1 / \omega_1) 
(\omega_1 t)^{- 10} / [\exp (a / \omega_1 t) - 1] \sin (\omega_1 t + \phi)$
with $a = 50$, $E_1 = 2 \times 10^{- 5}$ (100 kV/cm), $\omega_1 = 2 \times
10^{- 4}$ (1 THz) and $\phi = 0.3 \pi$, and the corresponding electric field
$E_1 (t) = - \partial_t A_1 (t)$ is shown in Fig. \ref{fig:scan-2d}(a).

When the THz field is absent, Fig. \ref{fig:scan-2d}(b) shows $| \tilde{d}
(\omega) |$ in the logarithmic scale with a typical plateau structure between
the 20th- and 80th-order. The harmonic yield dramatically decreases beyond the
cutoff around 80th-order, as depicted by the maximum kinetic energy of a
recolliding electron, $E_{\text{cutoff}} = I_p + 3.17 U_p$, where $U_p = E_0^2
/ 4 \omega_0^2$ is the ponderomotive energy of the electron.

When the THz field is present, the full scope of all-order $| \tilde{d}
(\omega) |$ in the logarithmic scale versus the time delay is presented in
Fig. \ref{fig:scan-2d}(c). To better observe the influence from the
accompanied THz field, the difference between $| \tilde{d} (\omega) |$ with
and without THz field, $\Delta | \tilde{d} (\omega) |$, is shown in Fig.
\ref{fig:scan-2d}(d) in the linear scale, with positive and negative values
indicated by distinct colors. A close scrutiny reveals most of the
nonvanishing $\Delta | \tilde{d} (\omega) |$ shown in (d) appear at even
orders. In the low-order region, the second-order harmonic follows the change
of THz intensity similar to the phenomenon utilized by TFISH, though the
applicability of the model in this low-order region is dubious. As the
harmonic order increases, the delay dependence of harmonics loses the
regularity and the distribution seems rather chaotic. When the harmonic order
increases up to the near-cutoff region, the delay-dependent harmonic yields,
however, again follow the time profile of $| E_1 (t) |$. Such a concurrence is
noteworthy, since it provides an alternative detection strategy to
characterize an arbitrary time-domain THz waveform.

\section{Analysis of near-cutoff even-order harmonics}\label{sec:analysis}

\subsection{Monochromatic light field}\label{subsec:mono}

Before the analysis of THz induced modulation on NCEHs, we retrospect the
simplest case where the HHG is induced by a monochromatic field of continuous
wave $E (t) = E_0 \cos (\omega_0 t)$ and the vector potential $A (t) = A_0
\sin (\omega_0 t)$ with $E_0 = - A_0 \omega_0$
{\cite{lewenstein_theory_1994}}. For concision, defining phases $\varphi_t =
\omega_0 t$ and $\varphi_{\tau} = \omega_0 \tau$, we have
$p_{\ensuremath{\operatorname{st}}} (\varphi_t, \varphi_{\tau}) = A_0 [\cos
\varphi_t - \cos (\varphi_t - \varphi_{\tau})] / \varphi_{\tau}$, and the
action in Eq. (\ref{eq:action-S}) reads
\begin{eqnarray}
  S_0 (\varphi_t, \varphi_{\tau}) & = & F_0 (\varphi_{\tau}) - \left(
  \frac{U_p}{\omega_0} \right) C_0 (\varphi_{\tau}) \cos (2 \varphi_t -
  \varphi_{\tau})  \label{eq:action-S0}
\end{eqnarray}
with the ponderomotive potential $U_p = E_0^2 / 4 \omega_0^2 = A_0^2 / 4$,
$F_0 (\varphi_{\tau}) = [(I_p + U_p) / \omega_0] \varphi_{\tau} - (2 U_p /
\omega_0)  (1 - \cos \varphi_{\tau}) / \varphi_{\tau}$, and
\begin{eqnarray}
  C_0 (\varphi_{\tau}) & = & \sin \varphi_{\tau} - \frac{4 \sin^2
  (\varphi_{\tau} / 2)}{\varphi_{\tau}} . 
\end{eqnarray}
Here, subscript "0" in $S_0$ is used to label the action without the
influence from the additional accompanied field, i.e., the THz field as will
be discussed in the subsequent sections. Applying the Anger-Jacobi expansion,
the exponential part $\exp (- \mathrm{i} S_0)$ with $S_0$ of Eq.
(\ref{eq:action-S0}) reads $\exp [- \mathrm{i} S_0 (\varphi_t,
\varphi_{\tau})] = \exp [- \mathrm{i} F_0 (\varphi_{\tau})]  \sum_{M = -
\infty}^{\infty} \mathrm{i}^M J_M (U_p C_0 (\varphi_{\tau}) / \omega_0) \exp
[\mathrm{i} M (\varphi_{\tau} - 2 \varphi_t)]$.

In Eq. (\ref{eq:d-sfa-generic}), the part including dipole matrix elements
$\mu^{\ast} [p_{\ensuremath{\operatorname{st}}} (\varphi_t, \varphi_{\tau}) +
A (\varphi_t)] \mu [p_{\ensuremath{\operatorname{st}}} (\varphi_t,
\varphi_{\tau}) + A (\varphi_t - \varphi_{\tau})] E_{} (\varphi_t -
\varphi_{\tau})$ can be represented by Fourier series $\sum_n b_n
(\varphi_{\tau}) \exp \left[ - \text{i}  (2 n + 1) \varphi_t \right]$ with
respect to $\varphi_t$. For simplicity, we assume the dipole moment of the
form $\mu (p) \sim \mathrm{i} p$, leading to mostly vanishing $b_n$ except for
ones with $2 n + 1 = \pm 1$ and $\pm 3$.

Applying the above expansions and substituting $n = K - M$, \ Eq.
(\ref{eq:d-sfa-generic}) eventually has the form
\begin{eqnarray}
  d_0 (t) & = & \sum_{K = - \infty}^{\infty} \tilde{d}_{0, 2 K + 1}
  \mathrm{e}^{- \mathrm{i} (2 K + 1) \varphi_t} .  \label{eq:d0-mono}
\end{eqnarray}
For $K \geqslant 0$, the coefficients for odd-order harmonics are given by
\begin{eqnarray}
  \tilde{d}_{0, 2 K + 1} & = & \mathrm{i} \int_0^{\infty} \mathrm{d} \tau
  \left( \frac{\pi}{\epsilon + \text{i} \tau / 2} \right)^{3 / 2}
  \mathrm{e}^{- \mathrm{i} F_0 (\varphi_{\tau})}  \nonumber\\
  &  & \times \sum_{M = 0}^{\infty} \mathrm{i}^M \mathrm{e}^{\mathrm{i} M
  \varphi_{\tau}} J_M \left( \frac{U_p}{\omega_0} C_0 (\varphi_{\tau}) \right)
  b_{K - M} (\varphi_{\tau}) \nonumber\\
  &  & + \text{c.c.}  \label{eq:d0-mono-coef}
\end{eqnarray}
while coefficients of all even-order harmonics vanish due to the restricted
value range of $n$ as long as the atomic potential is spherically symmetric.

\subsection{THz field induced NCEHs}\label{subsec:mono-and-THz}

An extra THz field of the same polarization direction, $A_1 (t)$, induces
even-order harmonics. In the followings, the detailed analysis of NCEHs is
presented to show their relations with the THz field. Defining $p_{\text{st},
i}$ the stationary momentum for a single field $i$ of vector potential $A_i
(t)$ with $i = 0$ and 1, the stationary momentum when both fields are present
satisfies $p_{\text{st}} = p_{\text{st}, 0} + p_{\text{st}, 1}$. Also, let
$S_i (t, \tau) = \frac{1}{2} \int_{t - \tau}^t \left[ p_{\text{st}, i} + A_i
(t) \right]^2 + I_p \tau$ be the single action for the $i$th field and note
that $\int_{t - \tau}^t A_i (t') \mathrm{d} t' = p_{\text{st}, i} \tau$, the
total action $S (t, \tau)$ relates to partial action $S_i (t, \tau)$ by
\begin{eqnarray}
  S (t, \tau) & = & S_0 (t, \tau) + [S_1 (t, \tau) - I_p \tau] \nonumber\\
  &  & - p_{\text{st}, 0} p_{\text{st}, 1} \tau + \int_{t - \tau}^t A_0 (t')
  A_1 (t') \mathrm{d} t' .  \label{eq:action-S-two-fields}
\end{eqnarray}
Given the vector potential of the additional THz field $A_1 (t) = A_1 \cos
(\omega_1 t + \phi)$ with $\omega_1 \ll \omega_0$ and $\phi$ an arbitrary
initial phase, we define the frequency ratio $\varepsilon = \omega_1 /
\omega_0 \ll 1$. Substituting $A_0 (t)$ and $A_1 (t)$ into Eq.
(\ref{eq:action-S-two-fields}), it is shown that the first-order correction
with regard to $\varepsilon$ arises completely from the cross term in Eq.
(\ref{eq:action-S-two-fields}) (i.e., the two terms on the second line), while
$S_1 (t, \tau) - I_p \tau$ merely has the contribution of $\mathcal{O}
(\varepsilon^2)$. Retaining terms of $S (t, \tau)$ only up to the first order
of $\varepsilon$, we have the yielding action $S (\varphi_t, \varphi_{\tau})
\simeq S_0 (\varphi_t, \varphi_{\tau}) + \Delta S (\varphi_t, \varphi_{\tau})$
with the THz-induced correction
\begin{eqnarray}
  \Delta S (\varphi_t, \varphi_{\tau}) & = & - p_{\text{st}, 0} p_{\text{st},
  1} \tau + \int_{t - \tau}^t A_0 (t') A_1 (t') \mathrm{d} t' \nonumber\\
  & \simeq & \varepsilon \frac{A_0 A_1}{\omega_0} \sin \phi C_1
  (\varphi_{\tau}) \cos \left( \varphi_t - \frac{\varphi_{\tau}}{2} \right) 
  \label{eq:action-dS}
\end{eqnarray}
and $C_1 (\varphi_{\tau}) = \varphi_{\tau} \cos (\varphi_{\tau} / 2) - 2 \sin
(\varphi_{\tau} / 2)$.

The action correction $\Delta S (\varphi_t, \varphi_{\tau})$ can also be
treated with Anger-Jacobi expansion, $\exp [- \mathrm{i} \Delta S (\varphi_t,
\varphi_{\tau})] = \sum_{N = - \infty}^{\infty} \mathrm{i}^N J_N (-
\varepsilon A_0 A_1 \sin \phi C_1 (\varphi_{\tau}) / \omega_0) \exp
[\mathrm{i} N (\varphi_{\tau} / 2 - \varphi_t)]$, resulting in the expansion
for the full action,
\begin{eqnarray}
  \mathrm{e}^{- \mathrm{i} S (\varphi_t, \varphi_{\tau})} & = & \mathrm{e}^{-
  \mathrm{i} F_0 (\varphi_{\tau})}  \sum_{M, N = - \infty}^{\infty}
  \mathrm{i}^{M + N} J_M \left( \frac{U_p}{\omega_0} C_0 (\varphi_{\tau})
  \right)  \nonumber\\
  &  & \times J_N \left( - \varepsilon \frac{A_0 A_1}{\omega_0} \sin \phi C_1
  (\varphi_{\tau}) \right) \nonumber\\
  &  & \times \mathrm{e}^{\mathrm{i} \left( M + \frac{N}{2} \right)
  \varphi_{\tau}} \mathrm{e}^{- \mathrm{i} (2 M + N) \varphi_t} . 
  \label{eq:exp-cw-thz}
\end{eqnarray}
Considering that the relatively small THz electric component $E_1 (t)$ is
negligible, the similar Fourier series expansion, \ $\mu^{\ast}
[p_{\ensuremath{\operatorname{st}}} (\varphi_t, \varphi_{\tau}) + A
(\varphi_t)] \mu [p_{\ensuremath{\operatorname{st}}} (\varphi_t,
\varphi_{\tau}) + A (\varphi_t - \varphi_{\tau})] E_{} (\varphi_t -
\varphi_{\tau}) = \sum_M b_M (\varphi_{\tau}) \exp \left[ - \text{i}  (2 M +
1) \varphi_t \right]$, can still be performed as in Sec. \ref{subsec:mono}
with $E (\varphi_t - \varphi_{\tau}) \simeq E_0 (\varphi_t - \varphi_{\tau})$.

\begin{figure}[h]
  \resizebox{226pt}{155pt}{\includegraphics[scale=1]{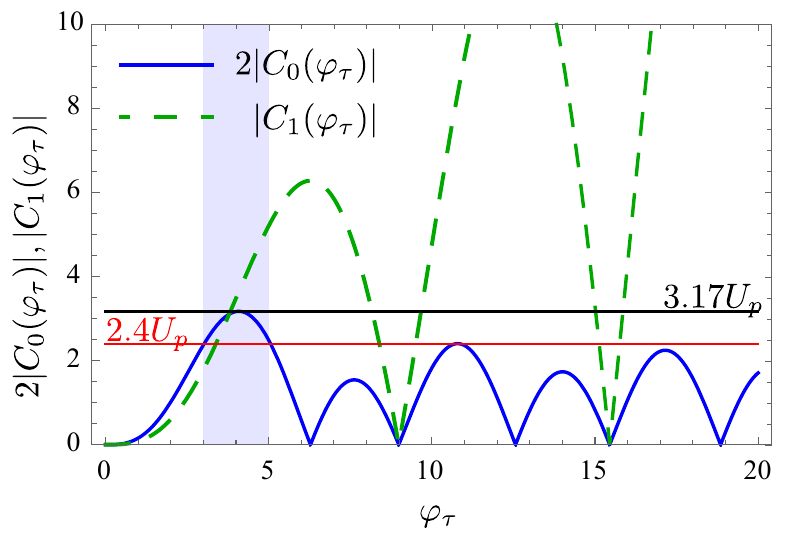}}
  \caption{\label{fig:C0-and-C1}The function $2 | C_0 (\varphi_{\tau}) |$
  (solid line) and $| C_1 (\varphi_{\tau}) |$ (dashed line) versus
  $\varphi_{\tau}$ of the return time. The maximum of the $2 | C_0
  (\varphi_{\tau}) |$ is associated to the maximum kinetic energy gain that
  corresponds to the cutoff energy of the HHG, as indicated by the black line
  for $3.17 U_p$. The shaded area (light blue) highlights $\varphi_{\tau}$
  contributing to the near-cutoff harmonic spectrum of the the energy between
  $2.4 U_p$ (red line) and $3.17 U_p$ (black line).}
\end{figure}

Substituting the above expansion and Eq. (\ref{eq:exp-cw-thz}) into Eq.
(\ref{eq:d-sfa-generic}), we obtain
\begin{eqnarray}
  d (t) & = & \text{i}  \sum_{n = - \infty}^{\infty} \int_0^{\infty}
  \mathrm{d} \tau \left( \frac{\pi}{\epsilon + \text{i} \tau / 2} \right)^{3 /
  2} b_n (\varphi_{\tau}) \mathrm{e}^{- \mathrm{i} F_0 (\varphi_{\tau})}
  \nonumber\\
  &  & \times \sum_{M, N = - \infty}^{\infty} \mathrm{i}^{M + N} J_M \left(
  \frac{U_p}{\omega_0} C_0 (\varphi_{\tau}) \right)  \nonumber\\
  &  & \times J_N \left( - \varepsilon \frac{A_0 A_1}{\omega_0} \sin \phi C_1
  (\varphi_{\tau}) \right) \nonumber\\
  &  & \times \mathrm{e}^{\mathrm{i} \left( M + \frac{N}{2} \right)
  \varphi_{\tau}} \mathrm{e}^{- \mathrm{i} [2 (M + n) + N + 1] \varphi_t} +
  \text{c.c.} .  \label{eq:d-THz}
\end{eqnarray}
The near-cutoff harmonics are usually featured with restricted range of
$\varphi_{\tau}$. As shown in Fig. \ref{fig:C0-and-C1}, the photon energy
within the range $2.4 U_p < 2 | C_0 (\varphi_{\tau}) | < 3.17 U_p$ requires
$\varphi_{\tau} \in [3, 5]$, corresponding to the first return of the released
electron (see also Fig. 1 of Ref. {\cite{lewenstein_theory_1994}}). Within the
range of $\varphi_{\tau}$ contributing to the near-cutoff harmonics, as
indicated by the shaded area in Fig. \ref{fig:C0-and-C1}, the globally
increasing function $C_1 (\varphi_{\tau})$ also remains low, resulting in the
whole argument of $J_N$ being small. Taking a typical 800 nm, $1 \times
10^{14}$ W/cm$^2$ femtosecond pulse and 1 THz, 1 MV/cm THz field for example,
the argument in $J_N (z)$ is roughly $| z | \approx 0.3$. Since $J_N (z)$
becomes exponentially small when $N > | z |$, only when $N = 0, \pm 1$ does
$J_N (z)$ significantly contribute, since we have the typical value $| z | <
1$. Using $J_N (z) \simeq (z / 2)^N / \Gamma (N + 1)$ when $z \rightarrow 0$
{\cite{abramowitz_handbook_1965}}, we find from Eq. (\ref{eq:d-THz}) that $d
(t) = d_0 (t) + d_1 (t) +\mathcal{O} (\varepsilon^2)$, where $d_0 (t)$, simply
given by Eq. (\ref{eq:d0-mono}) for odd-order harmonics in a monochromatic
field, stems from the contribution of $N = 0$ as $J_0 (z) = 1$. The odd-order
harmonics are barely influenced by the additional low-frequency field around
the near-cutoff region as long as the THz field remains relatively low to
fulfill $| z | < 1$ for $J_N (z)$. It is noteworthy that an extra correction
to the dipole moment, $d_1 (t)$, is induced by the THz field, corresponding to
$N = \pm 1$ as $J_{\pm 1} (z) = \pm z / 2$. In contrast to Eq.
(\ref{eq:d0-mono}), we have $d_1 (t)$ in the frequency domain,
\begin{eqnarray*}
  d_1 (t) & = & \sum_{K = - \infty}^{\infty} \tilde{d}_{1, 2 K} \mathrm{e}^{-
  \mathrm{i} (2 K) \varphi_t},
\end{eqnarray*}
containing only even-order harmonics, whose coefficients are given by
\begin{eqnarray}
  \tilde{d}_{1, 2 K} & = & \varepsilon \frac{A_0 A_1}{2 \omega_0} \sin \phi
  \int_0^{\infty} \mathrm{d} \tau \left( \frac{\pi}{\epsilon + \text{i} \tau /
  2} \right)^{3 / 2}  \nonumber\\
  &  & \times \mathrm{e}^{- \mathrm{i} F_0 (\varphi_{\tau})} C_1
  (\varphi_{\tau}) \sum_{M = 0}^{\infty} \mathrm{i}^M \mathrm{e}^{\mathrm{i} M
  \varphi_{\tau}} J_M \left( \frac{U_p}{\omega_0} C_0 (\varphi_{\tau}) \right)
  \nonumber\\
  &  & \times \left[ b_{K - M} (\varphi_{\tau}) \mathrm{e}^{-
  \frac{\mathrm{i}}{2} \varphi_{\tau}} + b_{K - M - 1} (\varphi_{\tau})
  \mathrm{e}^{\frac{\mathrm{i}}{2} \varphi_{\tau}} \right] \nonumber\\
  &  & + \text{c.c.} .  \label{eq:d1-mono-coef}
\end{eqnarray}
The prefactor of $\tilde{d}_{1, 2 K}$ thus suggests the proportionality to
$E_1$, showing a simple relation between NCEHs with the THz field. In
addition, the initial phase $\phi$ between the pair of continuous waves also
tunes the amplitudes of even-order harmonics, showing a sinusoidal dependence
of NCEHs on $\phi$.

\begin{figure}[h]
  \resizebox{238pt}{195pt}{\includegraphics[scale=1]{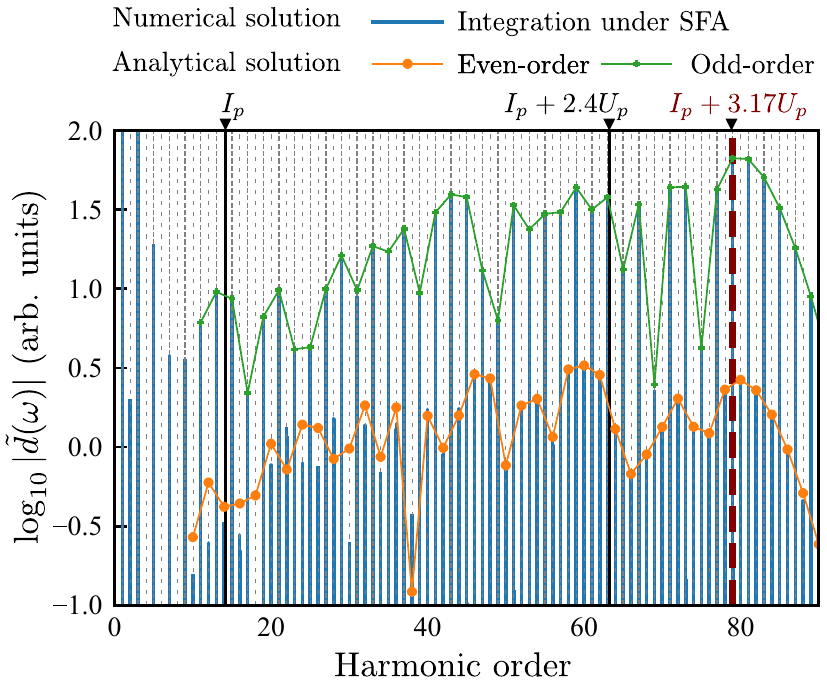}}
  \caption{\label{fig:hhg-cw}Emission of all-order harmonics under the laser
  field of continuous wave accompanied by a THz field. The laser parameters
  are given by $\omega_0 = 0.0353$ (1.3 $\mu$m) and $E_0 = 0.06$ (intensity
  $1.3 \times 10^{14}$ W/cm$^2$) for the infrared laser, and $\omega_1 = 3.53
  \times 10^{- 4}$ (2.3 THz) and $E_1 = 2 \times 10^{- 5}$ (100 kV/cm) for the
  THz field. The $| \tilde{d} (\omega) |$ (blue line), evaluated from the
  Fourier transform of $d (t)$ as the direct numerical integration of Eq.
  (\ref{eq:d-sfa-generic}), \ is compared with analytical formula, Eq.
  (\ref{eq:d0-mono-coef}) and Eq. (\ref{eq:d1-mono-coef}), for odd- (green)
  and even-order (orange) harmonics, respectively. The even-order harmonics
  are maximized with initial phase $\phi = \pi / 2$.}
\end{figure}

Fig. \ref{fig:hhg-cw} presents the comparison of harmonics $| \tilde{d}
(\omega) |$ from numerical integration of Eq. (\ref{eq:d-sfa-generic}) with
analytical formula, Eq. (\ref{eq:d0-mono-coef}) and Eq.
(\ref{eq:d1-mono-coef}), for odd- and even-order harmonics, respectively, when
initial phase $\phi = \pi / 2$ is chosen to maximize the yield of even-order
harmonics. The agreement between odd-order $| \tilde{d} (\omega) |$ evaluated
by Eq. (\ref{eq:d-sfa-generic}) with the analytical solution of Eq.
(\ref{eq:d0-mono-coef}) confirms the conclusion that the additional THz field
with the current laser parameters imposes no influences on odd-order harmonic
generation. On the other hand, the yields of even-order harmonics are
typically lower than their odd-order counterparts due to the small ratio of
frequencies $\varepsilon$ in the prefactor of Eq. (\ref{eq:d1-mono-coef}). The
derived solution of Eq. (\ref{eq:d1-mono-coef}) is also in excellent agreement
with the numerical result for NCEHs, showing the validity of the assumed
conditions that only $J_N (z)$ of $N = \pm 1$ contribute. Eq.
(\ref{eq:d1-mono-coef}) even works in a broader parametric range than expected
--- it correctly describes all even-order harmonics above 40th-order, which
corresponds to a much lower energy than that of the cutoff.

The above analysis establishes the basis for the generation of NCEHs. In the
following, the envelope effect for a more realistic laser pulse will be
presented to account for the time-resolving capacity of the femtosecond laser
pulse in THz detection.

\subsection{Effect of pulse envelope}\label{subsec:gaussian}

The envelope of a femtosecond laser pulse should be considered in practice. It
is expected that the temporal locality specified by the envelope plays an
essential role in determining the waveform of the THz field at exactly the
time of pulse center. In this section, we first discuss the envelope effect on
harmonics in the absence of the THz field.

Assuming the vector potential has a Gaussian-envelope, $A (t) = A_0 \exp [-
t^2 / (2 \sigma^2)] \sin (\omega_0 t)$, with the time center at $t = 0$ and
the pulse width $\sigma$, the excursion of the electron is given by $\alpha
(t) = - A_0 \exp [- (\omega_0 \sigma)^2 / 2] \sqrt{\pi / 2} \sigma
\mathrm{\ensuremath{\operatorname{Im}}} \left[
\mathrm{\ensuremath{\operatorname{erf}}} \left( t / \sqrt{2} \sigma +
\mathrm{i} \omega_0 \sigma / \sqrt{2} \right) \right]$ with the error function
$\text{erf} (z) = \left( 2 / \sqrt{\pi} \right)  \int_0^z \exp (- t^2)
\mathrm{d} t$. Substituting into Eq. (\ref{eq:action-S}), we find the action
\begin{eqnarray}
  S_0 (t, \tau) & = & I_p \tau + U_p \mathrm{e}^{- \left( \frac{t}{\sigma}
  \right)^2} \frac{\sqrt{\pi} }{2} \sigma G \left( \frac{t}{\sqrt{2} \sigma},
  \frac{\omega_0 \sigma}{\sqrt{2}}, \frac{\tau}{\sqrt{2} \sigma} \right) 
  \label{eq:action-S0-gaussian}
\end{eqnarray}
where
\begin{eqnarray*}
  G (x, y, \eta) & = & \mathrm{\ensuremath{\operatorname{Re}}}
  [\mathrm{e}^{\mathrm{i} 4 xy} g (x - \mathrm{i} y, \eta) - g (x, \eta)]\\
  &  & - \frac{\sqrt{2 \pi}}{\eta}  \left[
  \mathrm{\ensuremath{\operatorname{Im}}} \mathrm{e}^{\mathrm{i} 2 xy} g
  \left( \frac{x - \mathrm{i} y}{\sqrt{2}}, \frac{\eta}{\sqrt{2}} \right)
  \right]^2,
\end{eqnarray*}
and
\begin{eqnarray}
  g (z, \eta) & = & w \left( \mathrm{i} \sqrt{2} z \right) - \mathrm{e}^{4
  \eta (z - \eta / 2)} w \left( \mathrm{i} \sqrt{2} (z - \eta) \right) 
  \label{eq:func-g}
\end{eqnarray}
with $w (z)$ the Faddeeva function defined by $w (z) = \exp (- z^2)  [1 -
\mathrm{\ensuremath{\operatorname{erf}}} (- \mathrm{i} z)]$.

\begin{figure}[h]
  \resizebox{226pt}{155pt}{\includegraphics[scale=1]{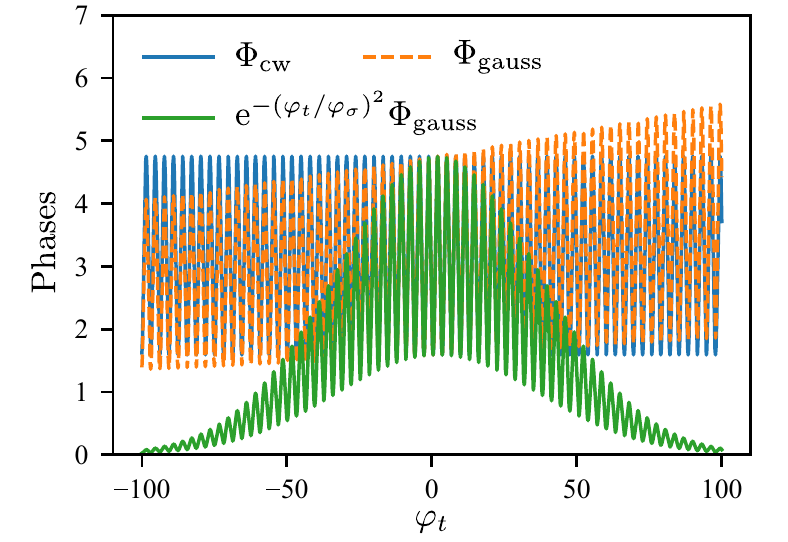}}
  \caption{\label{fig:Phi-comparison}Comparison of contributing phases in
  actions as a function of $\varphi_t$ between fields of continuous wave and
  of Gaussian envelope. Parameters $\varphi_{\sigma} = 50$ and $\varphi_{\tau}
  = 4$ are used which are typical for the generation of NCEHs.}
\end{figure}

In comparison with the action for a continuous wave, Eq. (\ref{eq:action-S0}),
which can be recast as $S_0 (\varphi_t, \varphi_{\tau}) = (I_p / \omega_0)
\varphi_{\tau} + (U_p / \omega_0) \Phi_{\text{cw}}$ with phase
$\Phi_{\text{cw}} = \varphi_{\tau} - 2 (1 - \cos \varphi_{\tau}) /
\varphi_{\tau} - C_0 (\varphi_{\tau}) \cos (2 \varphi_t - \varphi_{\tau})$,
action (\ref{eq:action-S0-gaussian}) for a Gaussian-enveloped pulse takes the
similar form, $S_0 (\varphi_t, \varphi_{\tau}) = (I_p / \omega_0)
\varphi_{\tau} + (U_p / \omega_0) \exp [- (\varphi_t / \varphi_{\sigma})^2]^{}
\Phi_{\text{gauss}}$, with $\Phi_{\text{cw}}$ replaced by a Gaussian-windowed
one $\exp [- (\varphi_t / \varphi_{\sigma})^2]^{} \Phi_{\text{gauss}}$ and
$\Phi_{\text{gauss}} = \left( \sqrt{\pi} / 2 \right) \varphi_{\sigma} G \left(
\varphi_t / \sqrt{2} \varphi_{\sigma}, \varphi_{\sigma} / \sqrt{2},
\varphi_{\tau} / \sqrt{2} \varphi_{\sigma} \right)$. Here, all notions of
phases $\varphi_t = \omega_0 t$ and $\varphi_{\tau} = \omega_0 \tau$ are still
used for consistency. Besides, we have introduced $\varphi_{\sigma} = \omega_0
\sigma$. A common factor of Gaussian $\exp [- (\varphi_t^{} /
\varphi_{\sigma})^2]$ in Eq. (\ref{eq:action-S0-gaussian}) specifies a
filtering window whose center coincides with that of the femtosecond pulse. A
pictorial analysis on the difference of actions between the continuous wave
and the Gaussian-enveloped pulse is presented in Fig.
\ref{fig:Phi-comparison}. The curves of both $\Phi_{\text{cw}}$ and
$\Phi_{\text{gauss}}$ contain the same dominant oscillating components versus
$\varphi_t$, proportional to $\cos (2 \varphi_t - \varphi_{\tau})$. Within the
region of interest, i.e., around the center of the Gaussian window, the
difference between $\Phi_{\text{gauss}}$ and $\Phi_{\text{cw}}$ is that the
oscillating amplitude of $\Phi_{\text{gauss}}$ increases with $\varphi_t$
while that of $\Phi_{\text{cw}}$ remains constant. Such an increase becomes
more significant with decreasing $\varphi_{\sigma}$. On the contrary, when
$\varphi_{\sigma}$ is sufficiently large, the amplitude of
$\Phi_{\text{gauss}}$ approaches $\Phi_{\text{cw}}$. That is, when the pulse
is infinitely long, i.e., $\varphi_{\sigma} \rightarrow \infty$, Eq.
(\ref{eq:action-S0-gaussian}) for a Gaussian envelope degenerates to Eq.
(\ref{eq:action-S0}), the action for a monochromatic continuous wave laser.

\

\begin{figure}[h]
  \resizebox{235pt}{155pt}{\includegraphics[scale=1]{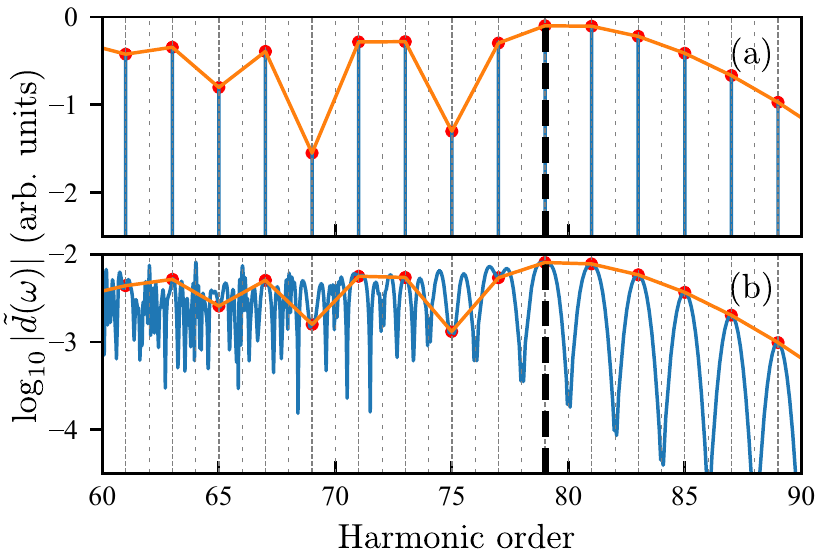}}
  \caption{\label{fig:hhg-envelope}Effect of pulse envelope on harmonic
  generation. The panels show the near-cutoff harmonics under (a) a continuous
  monochromatic laser of the same parameters used in Fig. \ref{fig:hhg-cw},
  and (b) a Gaussian-enveloped pulse of $\sigma = 1755$ (FWHM of 100 fs). The
  red marks label all harmonics of odd-orders, whose distribution is
  highlighted by orange curves. Black dashed lines indicate
  $E_{\text{cutoff}}$.}
\end{figure}

Fig. \ref{fig:hhg-envelope} shows the comparison of harmonics generations
between using a continuous wave and using Gaussian-enveloped femtosecond
pulse. With the same femtosecond laser parameters as considered for the
continuous wave ($\omega_0 = 0.0353$, $E_0 = 0.06$), the result in Fig.
\ref{fig:hhg-envelope}(a) is actually a zoom-in spectrum of Fig.
\ref{fig:hhg-cw} around the near-cutoff energy, while Fig.
\ref{fig:hhg-envelope}(b) shows the one with a Gaussian envelope of $\sigma =
100$ fs. The $| \tilde{d} (\omega) |$ at each odd-order, for either without or
with envelope effect, is highlighted by orange curve, showing the similar
distributions of odd-order harmonics. Action (\ref{eq:action-S0-gaussian})
modified by the finite pulse width, which has also been numerically examined,
however, contains extra frequency components, leading to multiple sidebands in
Fig. \ref{fig:hhg-envelope}(b) around the original odd-order harmonics. In the
next section the THz field combined with Gaussian-enveloped femtosecond pulse
is analyzed to show the theory behind the THz field reconstruction.

\subsection{THz induced NCEHs under Gaussian-enveloped
pulse}\label{subsec:gaussian-and-THz}

The above discussions allow for a straightforward extension to consider the
harmonic generation under a Gaussian-enveloped femtosecond laser pulse
accompanied by a THz field. Let $A (t) = A_0 \exp (- t^2 / 2 \sigma^2) \sin
(\omega t) + A_1 \cos (\omega_1 t + \phi)$ be the vector potential of the
combined fields. As presented in Sec. \ref{subsec:mono-and-THz}, the
corresponding action Eq. (\ref{eq:action-S-two-fields}) eventually takes the
form $S (t, \tau) \simeq S_0 (t, \tau) + \Delta S (t, \tau)$, where $S_0 (t,
\tau)$ is given by Eq. (\ref{eq:action-S0-gaussian}) for a Gaussian pulse as
presented in Sec. \ref{subsec:gaussian}, while the correction $\Delta S (t,
\tau)$ derives from the cross term $- p_{\text{st}, 0} p_{\text{st}, 1} \tau +
\int_{t - \tau}^t A_0 (t') A_1 (t') \mathrm{d} t'$. Similar to Sec.
\ref{subsec:mono-and-THz}, denoting the ratio between frequencies $\varepsilon
= \omega_1 / \omega_0$, solving $\Delta S (t, \tau)$ yields
\begin{eqnarray*}
  \Delta S (t, \tau) & = & - \frac{A_0 A_1}{2} \sqrt{\frac{\pi}{2}} \sigma
  \mathrm{e}^{- \left( \frac{t}{\sqrt{2} \sigma} \right)^2} 
  \mathrm{\ensuremath{\operatorname{Im}}} [\mathrm{e}^{\mathrm{i} \varphi_t} K
  (\varphi_t, \varphi_{\sigma}, \varphi_{\tau})],
\end{eqnarray*}
where
\begin{eqnarray*}
  K (\varphi_t, \varphi_{\sigma}, \varphi_{\tau}) & = & 2 \cos \left[
  \varepsilon \left( \varphi_t - \frac{\varphi_{\tau}}{2} \right) + \phi
  \right]  \mathrm{\ensuremath{\operatorname{sinc}} \left( \varepsilon
  \frac{\varphi_t}{2} \right)} g_0\\
  &  & - \mathrm{e}^{- \mathrm{i} (\varepsilon \varphi_t + \phi)} g_+ -
  \mathrm{e}^{+ \mathrm{i} (\varepsilon \varphi_t + \phi)} g_-
\end{eqnarray*}
with $g_0 \equiv g \left( z / \sqrt{2}, \eta / \sqrt{2} \right)$ and $g_{\pm}
\equiv g \left( z / \sqrt{2} \pm \mathrm{i} \varepsilon \varphi_{\sigma} / 2,
\eta / \sqrt{2} \right)$ defined by function $g (z, \eta)$ of Eq.
(\ref{eq:func-g}). Here, arguments $z$ and $\eta$ are dimensionless
compositions of time variables, $z = (\varphi_t / \varphi_{\sigma} -
\mathrm{i} \varphi_{\sigma}) / \sqrt{2}$ and $\eta = \varphi_{\tau} / \sqrt{2}
\varphi_{\sigma}$. With a small $\varepsilon$, the series expansion of $K
(\varphi_t, \varphi_{\sigma}, \varphi_{\tau})$ with respect to $\varepsilon$
up to the first order results in
\begin{eqnarray*}
  K (\varphi_t, \varphi_{\sigma}, \varphi_{\tau}) & \simeq & - 2 \sqrt{2}
  \varepsilon \mathrm{e}^{\mathrm{i} \varphi_t} \varphi_{\sigma} \sin \phi \Xi
  (z, \eta),
\end{eqnarray*}
where
\begin{eqnarray*}
  \Xi (z, \eta) & = & \left( z - \frac{\eta}{2} \right) g \left(
  \frac{z}{\sqrt{2}}, \frac{\eta}{\sqrt{2}} \right) - \frac{1 - \mathrm{e}^{2
  \eta \left( z - \frac{\eta}{2} \right)}}{\sqrt{\pi}} .
\end{eqnarray*}
Hence the action is given by
\begin{eqnarray}
  \Delta S (t, \tau) & \simeq & \sqrt{\pi} A_0 E_1 \mathrm{e}^{- \left(
  \frac{\varphi_t}{\sqrt{2} \varphi_{\sigma}} \right)^2} \sin \phi \sigma^2 
  \mathrm{\ensuremath{\operatorname{Im}}} [\mathrm{e}^{\mathrm{i} \varphi_t}
  \Xi (z, \eta)] .  \label{eq:action-dS-gaussian-THz-first-order}
\end{eqnarray}
With a typically large value of $\varphi_{\sigma}$ when the femtosecond laser
of several tens of optical cycles is used, the Faddeeva function $w (z) \simeq
\mathrm{i} z / \sqrt{\pi} (z^2 - 1 / 2)$ if $| z |^2 \geqslant 256$
{\cite{zaghloul_algorithm_2017}}. Using the approximation and explicitly
expanding the imaginary part in $\Delta S (t, \tau)$, the full expression can
be rearranged by trigonometric functions, whose coefficients, each as a
polynomial of time variables, can be further simplified by retaining only the
highest order of $\varphi_{\sigma}$. Eventually, we find
\begin{eqnarray*}
  \Delta S (t, \tau) & \simeq & \frac{A_0 E_1}{2 \omega_0^2} \mathrm{e}^{-
  \left( \frac{\varphi_t}{\sqrt{2} \varphi_{\sigma}} \right)^2} \sin \phi \{ 
  - \varphi_{\tau} \cos \varphi_t + 2 \sin \varphi_t\\
  &  & - \mathrm{e}^{\frac{\varphi_t \varphi_{\tau}}{\varphi_{\sigma}^2}}
  [\varphi_{\tau} \cos (\varphi_t - \varphi_{\tau}) + 2 \sin (\varphi_t -
  \varphi_{\tau}) \}  .
\end{eqnarray*}
When the pulse duration is sufficiently long, $\exp (\varphi_t \varphi_{\tau}
/ \varphi_{\sigma}^2) \simeq 1$, $\Delta S (t, \tau)$ approaches
\begin{eqnarray}
  \Delta S (t, \tau) & \simeq & - \frac{A_0 E_1}{\omega_0^2} \mathrm{e}^{-
  \frac{\varphi_t^2}{2 \varphi_{\sigma}^2}} \sin \phi C_1 (\varphi_{\tau})
  \cos \left( \varphi_t - \frac{\varphi_{\tau}}{2} \right), 
  \label{eq:action-dS-gaussian-THz}
\end{eqnarray}
recovering action (\ref{eq:action-dS}) under continuous waves as discussed in
Sec. \ref{subsec:mono-and-THz}, except for the presence of an extra prefactor
$\exp (- t^2 / 2 \sigma^2)$ that serves as a temporal window.

\begin{figure}[h]
  \resizebox{235pt}{167pt}{\includegraphics[scale=1]{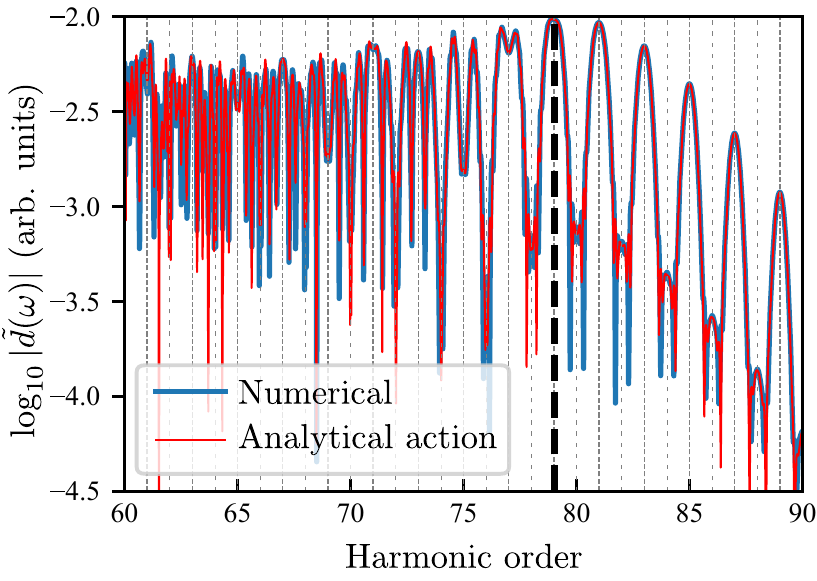}}
  \caption{\label{fig:hhg-gauss-thz-analytic}Comparison of $| \tilde{d}
  (\omega) |$ evaluated by direct numerical integration of Eq.
  (\ref{eq:d-sfa-generic}) (blue) and by using the derived action, $S = S_0 +
  \Delta S$, with $S_0$ and $\Delta S$ given by Eqs.
  (\ref{eq:action-S0-gaussian}) and (\ref{eq:action-dS-gaussian-THz}),
  respectively. The same parameters of femtosecond pulse as in Fig.
  \ref{fig:hhg-envelope} are used except for $\sigma = 2106$ (FWHM of 120 fs).
  The THz field is parametrized by $\omega_1 = 1 \times 10^{- 4}$, $E_1 = 2
  \times 10^{- 5}$ and $\phi = \pi / 2$.}
\end{figure}

\

Fig. \ref{fig:hhg-gauss-thz-analytic} shows the comparison of near-cutoff
harmonics evaluated by direct numerical integration of Eq.
(\ref{eq:d-sfa-generic}) with that obtained by applying the action of
analytical form, $S (t, \tau) = S_0 (t, \tau) + \Delta S (t, \tau)$, with $S_0
(t, \tau)$ and $\Delta S (t, \tau)$ given by Eqs.
(\ref{eq:action-S0-gaussian}) and (\ref{eq:action-dS-gaussian-THz}),
respectively. The comparison presents a rather good agreement, justifying the
analytically derived action with the assumed approximations. Comparing with
harmonics in Fig. \ref{fig:hhg-envelope}(b) without THz field, it is shown
that the odd-order harmonics are dominantly determined by $S_0 (t, \tau)$, as
those harmonics in both Fig. \ref{fig:hhg-envelope}(b) and Fig.
\ref{fig:hhg-gauss-thz-analytic} are almost the same, though odd-order
harmonic peaks in the latter are slightly sharper due to the use of longer
pulse width of the femtosecond laser. In Fig.
\ref{fig:hhg-gauss-thz-analytic}, however, NCEHs emerge, clearly indicating
that even-order harmonics originate from the THz-induced correction $\Delta S
(t, \tau)$.

From Eq. (\ref{eq:action-dS-gaussian-THz}), following the similar procedure of
analysis in Sec. \ref{subsec:mono-and-THz}, the reasoning behind the
generation of NCEHs in an enveloped laser pulse is straightforward. Within the
temporal window specified by the Gaussian envelope, the strength of NCEHs is
approximately proportional to THz field strength exactly at the center of the
envelope. In other words, under the influence of the THz field that induces
even-order harmonics, the femtosecond pulse with a filtering temporal window
maps the instantaneous strength of THz field onto that of NCEHs, allowing for
a complete characterization of the THz time-domain spectrum with the
femtosecond pulse scanning over the THz field.

\begin{figure}[h]
  \resizebox{235pt}{181pt}{\includegraphics[scale=1]{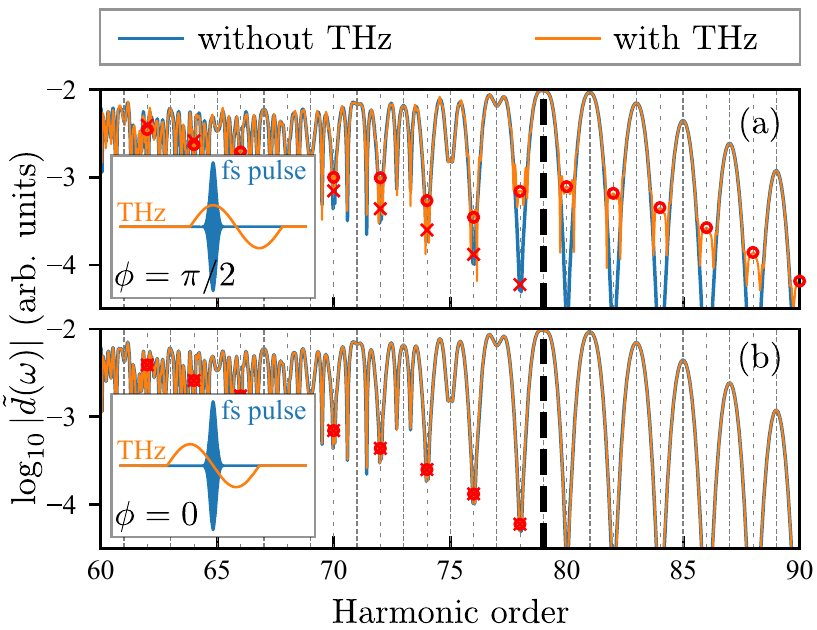}}
  \caption{\label{fig:hhg-thz-phase}Near cut-off harmonics without THz field
  (blue) and with THz field (orange) when (a) $\phi = \pi / 2$ and (b) $\phi =
  \pi$. The inset of each panel shows the time center of femtosecond pulse
  (blue) relative to that of the THz field (orange) for different $\phi$.
  Marks "$\circ$" (``$\times$'') label the even-order harmonics with
  (without) the THz field. The position of $E_{\text{cutoff}}$ is indicated by
  the black dashed line. The same parameters as in Fig.
  \ref{fig:hhg-gauss-thz-analytic} are used.}
\end{figure}

Fig. \ref{fig:hhg-thz-phase} shows the dependence of NCEHs on initial phase
$\phi$, or equivalently, the pulse center of the femtosecond laser relative to
the electric component of the THz field. When $\phi = \pi / 2$, factor $\sin
\phi = 1$ in Eq. (\ref{eq:action-dS-gaussian-THz}) maximizes the coefficient
of NCEHs as analyzed in Eq. (\ref{eq:d1-mono-coef}). As shown in Fig.
\ref{fig:hhg-thz-phase}(a), the even-order harmonics under a THz field is
significantly higher than its counterpart without the THz field, and the
amplitude relative to their adjacent odd-order harmonics becomes even more
significant when the order approaching the cutoff. On the contrary, in Fig.
\ref{fig:hhg-thz-phase}(b), when $\phi = 0$, the even-order harmonics vanish
and the harmonic distribution in the presence of THz field is exactly the same
as the one without the THz field. Except for even-order harmonics, the
harmonics of other energies are almost identical between panels (a) and (b),
showing they are barely influenced by the THz field. As indicated by insets of
Fig. \ref{fig:hhg-thz-phase}, at $\phi = \pi / 2$ ($\phi = 0$), the center of
the femtosecond pulse is at the maximum (the zero-point) of the THz electric
field. The coincidence of the NCEHs yields with the THz electric field shows
the feasibility to reconstruct the latter using the former.

\begin{figure}[h]
  \resizebox{240pt}{240pt}{\includegraphics[scale=1]{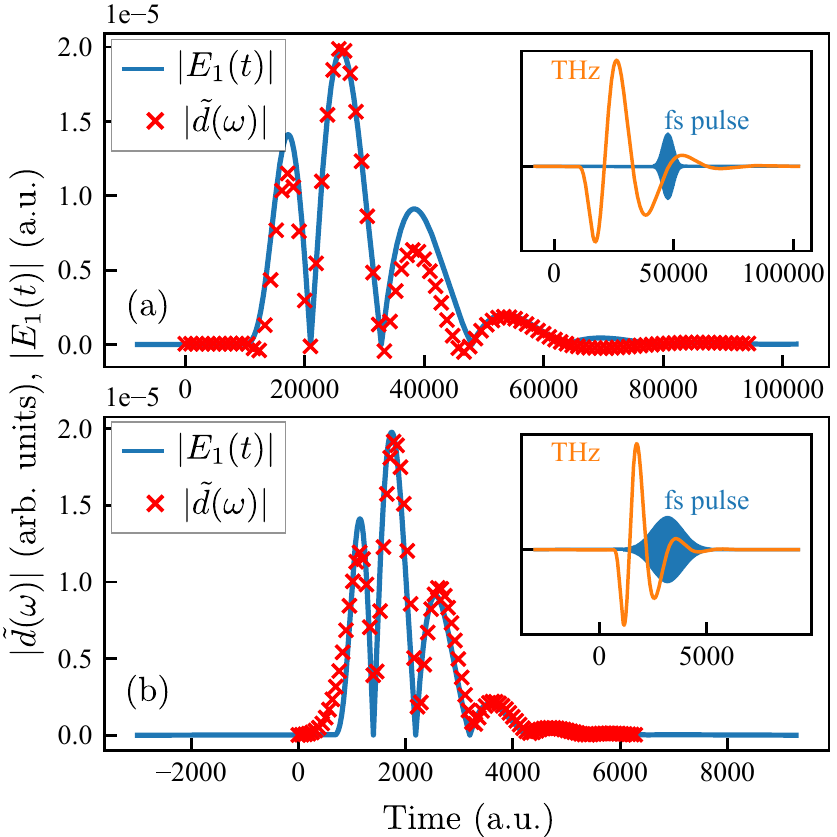}}
  \caption{\label{fig:reconstruct}Reconstruction of time-domain spectrum of
  THz waves. (a) The parameters are the same as used in Fig.
  \ref{fig:scan-2d}. $\omega_0 = 0.0353$ (1.3 $\mu$m), $E_0 = 0.06$ ($I = 1.3
  \times 10^{14}$ W/cm$^2$), $\sigma = 2106$ (FWHM of 120 fs), $\omega_1 = 2
  \times 10^{- 4}$ (1 THz), and $E_1 = 2 \times 10^{- 5}$ (100 kV/cm). (b) The
  reconstruction for THz wave of higher frequency with $\omega_0 = 0.1139$
  (400 nm), $E_0 = 0.06$ ($I = 1.3 \times 10^{14}$ W/cm$^2$), $\sigma = 702$
  (FWHM of 40 fs), $\omega_1 = 0.003$ (20 THz), and $E_1 = 2 \times 10^{- 5}$
  (100 kV/cm). Insets show the temporal profiles $E_0 (t)$ and $E_1 (t)$ of
  the femtosecond pulse and the THz field, respectively.}
\end{figure}

In Fig. \ref{fig:reconstruct}, the reconstruction of THz field from the NCEHs
is demonstrated by examples. Changing the time delay between the femtosecond
pulse and the THz field, the even-order harmonic nearest to the cutoff on the
lower energy side is retrieved and compared with $| E_1 (t) |$ of the THz
field. Using the same parameters of fields as mentioned above, we present the
reconstruction with the 78th-order NCEH in Fig. \ref{fig:reconstruct}(a),
which in general shows a good agreement of $| d (\omega) |$ with $| E_1 (t)
|$. Another example to detect the THz field of higher frequency is presented
in (b) to show the universality of the scheme. In order to resolve the THz
field of 20 THz, a femtosecond pulse of higher frequency is required to ensure
a low ratio $\varepsilon = \omega_1 / \omega_0$. Using a 400 nm laser pulse
with $\varepsilon \simeq 0.03$ and reducing the pulse width to 40 fs, the
generated 6th-order harmonic can also be used to reveal the time-domain THz
wave. The successful reconstruction of THz wave of short-time scale suggests
the possibility of THz broadband detection under the aid of the NCEHs
measurement.

The applicability of the reconstruction scheme is closely related to the
approximations applied for the analysis in previous sections. From the
temporal perspective, $\varepsilon = \omega_1 / \omega_0 \ll 1$ is a must,
indicating that the characterization of THz field of high frequency needs high
frequency femtosecond pulse. Moreover, the approximation of Faddeeva function
to solve Eq. (\ref{eq:action-dS-gaussian-THz-first-order}) requires
$(\varphi_t^2 / \varphi_{\sigma}^2 + \varphi_{\sigma}^2) / 2 > 256$,
necessitating $\varphi_{\sigma} > 23$, suggesting a femtosecond pulse should
contain as least 9 cycles within its FWHM. The range of $\varphi_{\sigma}$
also naturally satisfies both conditions that $\eta = \varphi_{\tau} /
\varphi_{\sigma} \ll 1$ and $\exp (\varphi_t \varphi_{\tau} /
\varphi_{\sigma}^2) \simeq 1$ to derive (\ref{eq:action-dS-gaussian-THz}). In
general, the scheme favors the use of large $\varphi_{\sigma}$, which also
helps suppress the sideband caused by the finite pulse width. Nevertheless, a
smaller $\varphi_{\sigma}$ allows for a better time resolution of the waveform
reconstruction. Therefore, a balance inbetween should be considered for the
choice of $\varphi_{\sigma}$, which also depends on the frequency range of THz
field to detect.

In addition, the choice of laser parameters, including field amplitudes $E_0$,
$E_1$ and frequency $\omega_0$, is critical to the applicability of the
reconstruction scheme. The small value of the argument of the Bessel function
in Eq. (\ref{eq:d-THz}) imposes the condition $| \varepsilon (A_0 A_1 /
\omega_0) C_1 (\varphi_{\tau}) | = | (E_0 E_1 / \omega_0^3) C_1
(\varphi_{\tau}) | < 1$. As shown in Fig. \ref{fig:Phi-comparison}, the value
of $| C_1 (\varphi_{\tau}) | \in [1.7, 5.2]$ when $\varphi_{\tau} \in [3, 5]$
for near-cutoff harmonics allows for an estimation of the loosely restricting
criterion, $E_0 E_1 / \omega_0^3 < 0.2$. Moreover, neglecting the THz field
$E_1 (t)$ in the prefactor of dipole matrix elements, $\mu^{\ast}
[p_{\ensuremath{\operatorname{st}}} (\varphi_t, \varphi_{\tau}) + A
(\varphi_t)] \mu [p_{\ensuremath{\operatorname{st}}} (\varphi_t,
\varphi_{\tau}) + A (\varphi_t - \varphi_{\tau})] E_{} (\varphi_t -
\varphi_{\tau})$, requires $E_1 \ll E_0$. Both conditions indicate an upper
limit for $E_1$. That is, the detected THz field in this work is not supposed
to be overwhelmingly intense, otherwise the approximation of the Bessel
function in Eq. (\ref{eq:d-THz}) breaks down, resulting in nonvanishing
high-order components that contribute to other complicated effects accompanied
by a strong low-frequency field, e.g., the field-induced multi-plateau
structure. Since the theory works in the tunneling regime, $I_p \leqslant 2
U_p$, the femtosecond laser also satisfies $E_0^2 > 2 \omega_0^2 I_p$.

Besides the conditions required to justify the reconstruction scheme, we also
need to take the finite signal-to-noise ratio into account. The even-order
harmonics should be large enough to observe. Assuming the amplitude of
even-order harmonics to the adjacent odd-order one is no less than one
percent, we may impose an extra condition with the coefficients of harmonics,
$\varepsilon A_0 A_1 / 2 \omega_0 = E_0 E_1 / 2 \omega_0^3 > 10^{- 2}$,
yielding $E_0 E_1 / \omega_0^3 > 0.02$. In contrast to the condition specified
by the approximation of Bessel function, it indicates the field amplitudes
should be large enough to generate even-order harmonics.

\begin{figure}[h]
  \includegraphics[scale=1]{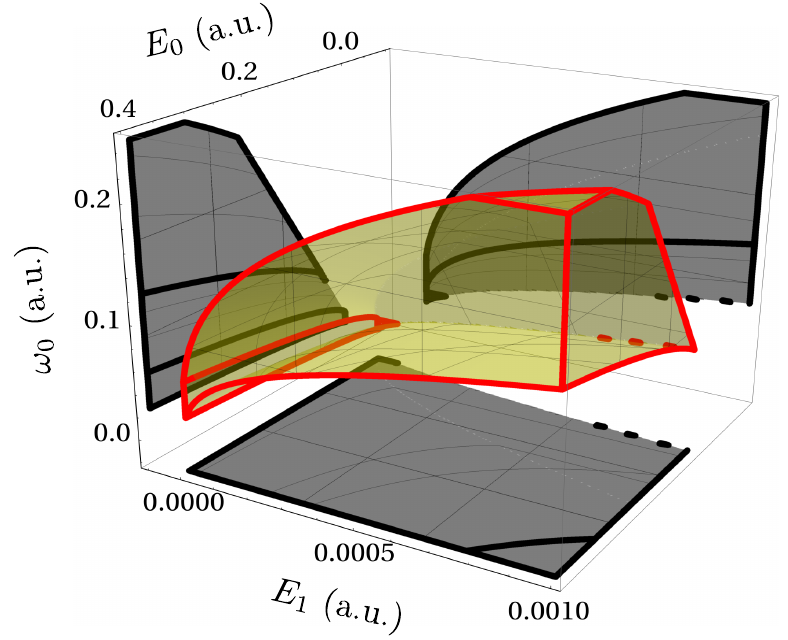}
  \caption{\label{fig:parameters}Available parametric range for THz
  reconstruction with NCEHs with regard to frequency $\omega_0$ and field
  amplitudes $E_0$ and $E_1$. The detailed conditions are specified in the
  main text.}
\end{figure}

All above conditions for the reconstruction scheme can be pictorially
illustrated in the parametric space as shown in Fig. \ref{fig:parameters},
where the appropriate parametric range is highlighted. When $E_1$ of the THz
field is low, a femtosecond pulse of longer wavelength avails the measurement;
on the contrary, the THz field of increasing $E_1$ requires a femtosecond
pulse of higher $\omega_0$, whose optional frequency range also becomes
broader. Concerning the relation $\varepsilon = \omega_1 / \omega_0 \ll 1$,
the accessible frequency of the THz field for waveform reconstruction thus
depends on field amplitudes. Especially both $E_0$ and $E_1$ being high favors
the use of higher $\omega_0$, allowing for the detection of THz field of
higher $\omega_1$.

\section{Summary and conclusion}

The harmonic generation by a half-wave symmetric driving laser that interacts
with isotropic media has long been known to yield odd-order harmonics only
{\cite{ben-tal_effect_1993}}. The emergence of even-order harmonics usually
attributes to certain broken symmetries {\cite{neufeld_floquet_2019}}, e.g.,
the THz field induced broken symmetry in this work. Here, the even-order
harmonic generation near the cutoff is found to have a particularly
synchronous relation with the THz electric field. The analytical derivation
with perturbative expansion shows the NCEHs originate from the first-order
correction with regard to the ratio between frequencies of the THz field and
that of the femtosecond laser pulse. The linear relation between the NCEH
amplitude and the THz electric field derives from an approximation of the
Bessel function, which can be fulfilled by the specific range of return time
corresponding to the near-cutoff energy region.

The direct mapping from the THz field to NCEHs thus provides an alternative
conceptually simple approach to reconstruct time-domain THz wave from NCEHs.
The proposal to measure NCEHs as a function of the time delay between the
femtosecond laser pulse and the THz field has been numerically verified,
showing the applicability of the method for the broadband THz detection. The
analytical derivations also help identify the parametric region for such
applications, indicating high-frequency THz field characterization should
require higher laser intensity. The encoding of the time-domain information of
THz wave into NCEHs may inspire new routes towards the realization of coherent
detection in broad spectral range.

\begin{acknowledgments}
  This work is supported by the National Natural Science Foundation of China
  (Grants No. 11874368, No. 11827806 and No. 61675213).
\end{acknowledgments}

\bibliographystyle{apsrev4-1}
\bibliography{main}

\end{document}